# Oxygen Vacancies on SrO-terminated SrTiO$_3$(001) Surfaces studied by Scanning Tunneling Spectroscopy

Wattaka Sitaputra[1], Nikhil Sivadas[2], Marek Skowronski[1], Di Xiao[2], Randall M. Feenstra[2]

[1]*Dept. Materials Science and Engineering, Carnegie Mellon University, Pittsburgh, PA 15213*
[2]*Dept. Physics, Carnegie Mellon University, Pittsburgh, PA 15213*

The electronic structure of SrTiO$_3$(001) surfaces was studied using scanning tunneling spectroscopy and density-functional theory. With high dynamic range measurements, an in-gap transition level was observed on SrO-terminated surfaces, at 2.7 eV above the valence band maximum. The density of centers responsible for this level was found to increase with surface segregation of oxygen vacancies and decrease with exposure to molecular oxygen. Based on these finding, the level is attributed to surface O vacancies. A level at a similar energy is predicted theoretically on SrO-terminated surfaces. For TiO$_2$-terminated surfaces, no discrete in-gap state was observed, although one is predicted theoretically. This lack of signal is believed to be due to the nature of defect wavefunction involved, as well as the possible influence of transport limitations in the tunneling spectroscopy measurements.

## I. INTRODUCTION

Complex oxide systems are presently of great interest, in part because *interfaces* between such materials exhibit properties different from the constituents [1–3]. LaAlO$_3$/SrTiO$_3$ heterostructure is a typical example of such a system. It is known to exhibit 2-dimensional electron gas (2DEG) and ferromagnetism at the interface for LaAlO$_3$ thickness of at least 4 unit cells [4–8]. However, even after a decade since its discovery, the actual driving force behind 2DEG formation is still not well understood; the major sources of electron doping at the interface are being debated as possibly due to polar catastrophe [9–12], doping by oxygen vacancies [4,13,14] and cation intermixing [15–17]. In the related, simpler system of a SrTiO$_3$(001) surface, the formation of 2DEG is largely accepted as due to surface oxygen vacancies [18–20]. It was found that, by exposing a low temperature, vacuum-cleaved surface of SrTiO$_3$(001) to strong ultraviolet light, a defect level at 1.3 eV below the Fermi level was created together with the formation of 2DEG [18]. Intriguingly, this well-known oxygen vacancy state [18,21–25] lies too deep below the conduction band to provide carriers and form the 2DEG.

      In this study, we have used scanning tunneling spectroscopy (STS) to selectively probe different terminations of the SrTiO$_3$(001) surface, i.e. SrO and TiO$_2$ terminations, and we study their electronic structures that arise from the presence of surface oxygen vacancies. Surfaces are prepared both by cleaving in ultra-high-vacuum (UHV) and by growth by molecular-beam epitaxy (MBE). We find that an in-gap level is produced by vacancies residing on a SrO-terminated surface. This result is the first direct observation of an in-gap transition level for a SrO-terminated surface. The position of this level with respect to the Fermi energy was found to vary with a roughness of the surface, signifying the presence of coexisting disorder-induced surface states. On the other hand, no such level is observed for a vacancy on a surface TiO$_2$

plane. To support our experimental observations and their interpretation, we employ first-principles predictions of the oxygen vacancy electronic structure, using the LSDA + U method. The results show different positions of the transition levels for different terminating planes. For the SrO termination we predict a donor level, i.e. (0/+) transition level, in approximate agreement with our experimental observations. For the TiO$_2$ termination we predict a donor level that is resonant with the conduction band, in agreement with prior theory and experiment [18,19,26]. For both terminations we also predict in-gap levels for the double donor, i.e. (+/++) transition level. These levels are not observed experimentally, although we argue that their absence occurs due to limited sensitivity of the measurement (either because of limitations in carrier transport, or limitations in surface wavefunction extent, or both).

## II. EXPERIMENTAL DETAILS

In this work, both cleaved and homoepitaxially grown surfaces were studied. The cleaved surfaces of SrTiO$_3$(001) were prepared by fracturing 0.05 wt% Nb-doped SrTiO$_3$(001) substrates along (001) plane in ultra-high vacuum (UHV) at room temperature. Prior to cleaving, our samples were sputter coated with 100 nm of titanium on both polished surfaces. This was done in order to ensure uniform power dissipation through the sample when a bias is applied between each surface during resistive heating. In addition, the titanium also reduces the sample, creating oxygen vacancies during the outgassing step which was performed by annealing at 700-800˚C for 5 minutes [27].

For the homoepitaxially grown surfaces, MBE was performed by co-depositing titanium and strontium onto 0.01 wt% Nb-doped SrTiO$_3$(001) substrates, which were prepared to have TiO$_2$-terminated surface using the Arkansas method [28]. Substrate temperature, deposition rate and partial pressure of a molecular oxygen were kept at 750ºC, 20 seconds per monolayer and 10$^{-6}$ Torr, respectively. The thickness of the films was kept below 15 unit cells for all samples. A Ta susceptor and a Pt (50 nm)/Ti (20 nm) back coating were used not only for absorbing the 808 nm laser for substrate heating but also for aiding a reduction of the sample. Formation of oxygen vacancies was ensured by stopping the oxygen supply when the substrate temperature reached 600ºC during the cooling down after the deposition. At the end of the process, the samples were also visually inspected and found to appear darker, reaffirming the presence of oxygen vacancies.

Scanning tunneling microscopy (STM) and spectroscopy were performed at room temperature with Pt/Ir tips. Tunnel currents in the range 0.1-0.5 nA and sample biases in the range 1.5 - 3.0 V were used for acquisition of topographic (i.e. constant current) images and conductance maps. A lock-in technique was used to obtain differential tunneling conductance (dI/dV) spectra with oscillation frequency and rms modulation amplitude of 1 kHz and 50 mV, respectively, for cleaved surfaces, and 15 kHz and 25 mV, respectively, for MBE-grown surfaces.

An important technical aspect of our STS measurements is our method of obtaining relatively high dynamic range, which is necessary in order to observe mid-gap states [29–31]. Figure 1(a) presents a typical spectrum acquired from the cleaved SrTiO$_3$ surface, using a fixed tip-sample separation. Here, as in the work of Guisinger et al. [32], mid-gap states are not observed in the spectra since the acquisition has a limited dynamic range and the data is plotted on a linear conductance scale. Even if it were to be plotted in logarithmic scale, the signal to noise is relatively low (only 1-2 orders of magnitude of the data are above the noise level), so that, again, mid-gap features would not be seen. In order to obtain a higher sensitivity, the tip-sample separation is varied as a function of applied bias, i.e. the tip moves closer to the sample as the magnitude of the bias decreases [33]. Following the measurement, the exponential increase in conductance due to the variation in tip-sample separation is normalized by multiplying the data by a factor of $e^{-2\kappa Z(V)}$ where $\kappa$ is an experimentally-determined decay constant and $Z(V)$ is the bias-dependent change in tip-sample separation. This procedure yields spectra such as that shown in Fig. 1(b), with 3-4 orders of magnitude of dynamic range. $Z(V)$ is varied according to $a|V|$ where the constant $a$ is typically 1.5 Å/V, and the experimentally determined value of $\kappa$ is typically 4.5 nm$^{-1}$. (This experimental value is less than an ideal $\kappa$ of 10 nm$^{-1}$, due to effects such as residual surface charging [33,34].) We emphasize that this normalization only affects the *extent* of the conductance axis in Fig. 1(b), but doesn't affect any detailed structure within the spectrum. With the improved sensitivity, in-gap states are clearly observed in Fig 1(b). These states arise predominantly from surface disorder that occurs on the cleaved TiO$_2$-terminated surface areas, as discussed in Section III(A).

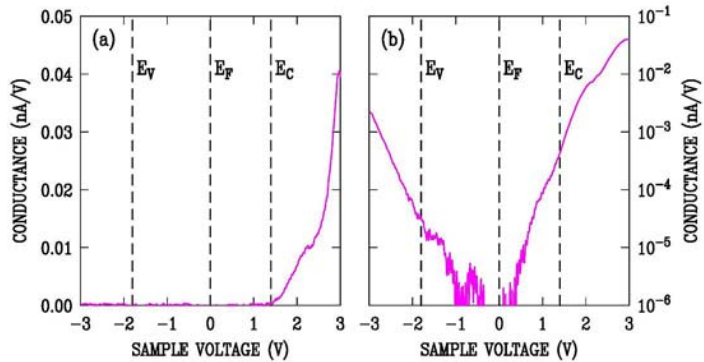

**FIG. 1** (Color online) (a) Typical conductance spectra obtained with fixed tip-sample separation, plotted on a linear scale. (b) Conductance versus voltage spectrum obtained with varying tip-sample separation and hence with higher sensitivity. Conduction and valence band edges are approximately located based on the spectrum in (b), with the same positions shown in (a). Note: these spectra were obtained by averaging results from across the surface, including both SrO and TiO$_2$ termination.

## III. EXPERIMENTAL RESULTS AND DISCUSSION

### A. Cleaved surfaces

Fracture of the samples produces surface morphologies with varying roughness (step density), consistent with the prior results of Guisinger *et al*. [32] Figure 2 shows STM results for our typical cleaved surfaces of SrTiO$_3$(001). Upon fracture at room temperature, conductance stripes arising from alternating SrO and TiO$_2$ terminated-terraces are observed as shown in Fig. 2(b) and (c). Additionally, the two terminations can reveal varying amounts of surface disorder which, for Fig. 2, takes the form of greater roughness on the TiO$_2$-terminated surface. The two terminations can be reliably distinguished if one knows the difference in local density of states between the terminations at a particular sample bias [32,35,36]. A bright stripe seen in a conductance map acquired with a sample bias of +3.0 V generally signifies SrO termination while a dark stripe signifies TiO$_2$ termination. However, for lower sample biases as often used in our experiments, this contrast can be reversed due to a higher local density of states at the conduction band (CB) edge for the TiO$_2$-terminated terraces (as reported by Guisinger *et al.* [32] and also revealed in detail in the spectra below).

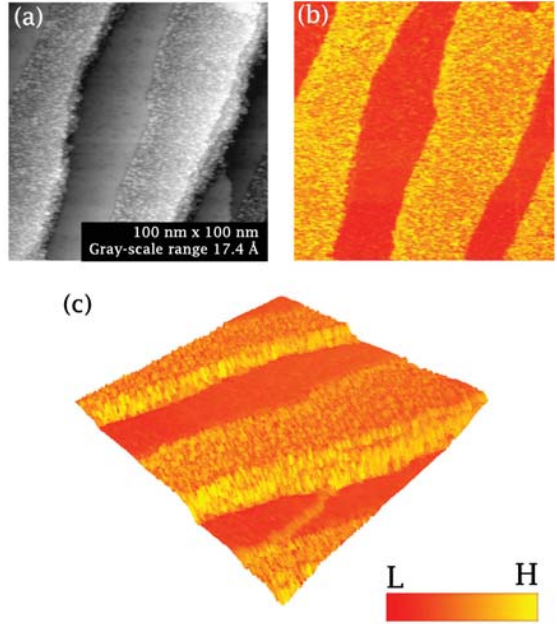

**FIG. 2** (Color online)(a) Topographic image, (b) conductance map, and (c) perspective overlay between topographic image and conductance map (c) of the cleaved SrTiO$_3$(001) surface obtained with a sample bias of +1.5 V and tunnel current of 0.1 nA. At this bias, bright conductance stripes occur for TiO$_2$ termination while dark stripes occur for SrO termination.

Figure 3 presents STS results from the two surface terminations. Each curve in these plots represents an average of 6 – 12 spectra acquired over the specified termination. Figure 3(a) shows a comparison of results of as-cleaved surfaces, while Fig. 3(b) compares the spectra after an additional preparation step consisting of 10 minutes of moderate-temperature annealing (260 - 360ºC). For the as-cleaved surfaces, on the TiO$_2$-terminated terraces, only a typical onset associated with the bulk SrTiO$_3$ CB was observed, with shoulder located about 2.0 eV above the Fermi level. On the SrO-terminated terraces, this CB onset is shifted upwards by about 0.25 eV; it is this characteristic shift that allows us to distinguish the two surface terminations. An onset for the valence band (VB) is also seen (at negative voltages) on the SrO-terminated terraces, and additionally, a broad weak feature centered near 1.3 eV above the Fermi level is observed.

A large increase in the intensity of this in-gap feature on the SrO termination was observed after annealing, as shown in Fig. 3(b). We attribute this feature to surface segregation of bulk oxygen vacancies, generated during outgassing step. A drastic decrease in intensity of this peak was observed (Fig. 3(c)) upon exposure to 10 Langmuir of molecular oxygen. It is also worth noting that our annealing temperature of 260 – 360°C is not sufficient to *create* a significant number of oxygen vacancies [37,38]. Instead, the high concentration of oxygen vacancies produced during the outgassing step becomes supersaturated and they segregate to the surface. As the result, the system free energy is lowered due to lower enthalpy of surface vacancies [39] (even though the entropic contribution, -T∆S, increases). In such a case, the moderate temperature annealing merely serves to increase the rate at which oxygen vacancies diffuse toward the surface.

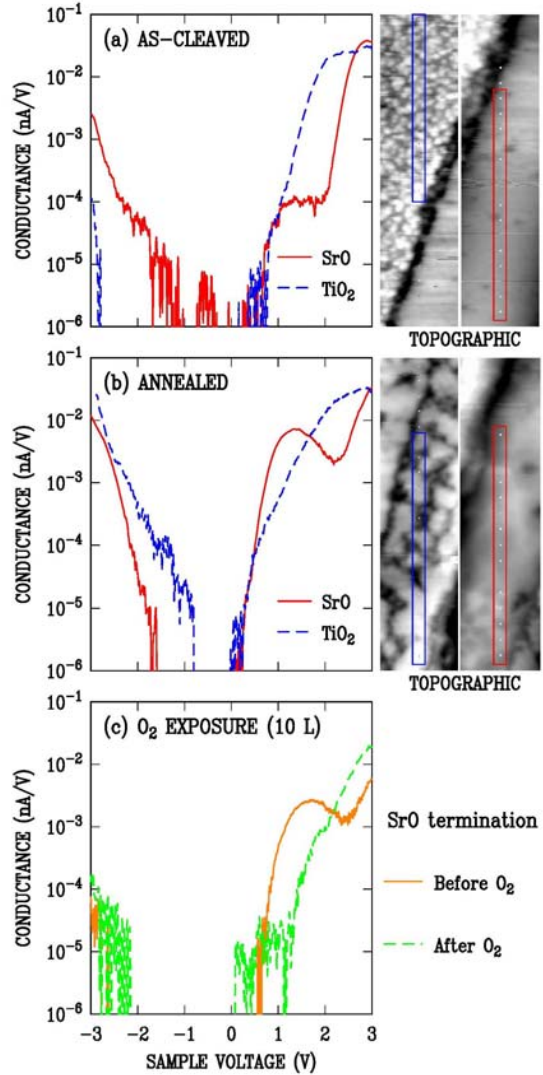

**FIG. 3** (Color online) Average conductance spectra for SrO and TiO$_2$ termination acquired (a) before and (b) after segregation of surface oxygen vacancies by moderate-temperature annealing. Two rectangular topographic images (130 nm x 500 nm) shown on the right of each plot illustrate the areas on which the spectra were averaged over, i.e. left-hand image (blue box) for TiO$_2$ termination and right-hand image (red box) for SrO termination. The topographic images were acquired with a sample bias of +3.0 V. (c) Comparison between conductance spectra from the annealed SrO-terminated surfaces before and after 10 L of molecular oxygen exposure. The sample voltage corresponds to the energy of a state relative to the Fermi level (0 V in the spectra).

By extrapolating the monotonically increasing portions of the spectra associated with the VB and CB edges in Fig. 3(b) for the SrO termination toward a noise level (10$^{-6}$ nA/V), and overlaying the spectrum of Fig. 3(a) to help define the CB edge, we determine the locations of the VB and CB band edges as -1.6 V and +1.7 V, respectively. Their difference is consistent with the known SrTiO$_3$ bandgap of 3.2 eV, indicating the tip-induced band bending is relatively small on this area of the surface [40]. The location of the in-gap feature on the SrO terraces, associated with oxygen vacancies, is thus found to be centered at 2.9 V above the VB maximum. We note that this in-gap feature was not observed in previous reports,[44] most likely because the measurement sensitivity was insufficient.

In addition to the peak discussed above, annealing also increased the signal at negative voltages on TiO$_2$ termination. We attribute this conductance tail extending out from the VB to in-gap states induced by increased disorder. As shown in the side images of Figs. 3(a) and (b), the annealing leads to restructuring of the TiO$_2$ surface plane; the surface which is initially rough on an atomic scale develops topography with more distinct terraces separated by steps, as will be discussed in more detail elsewhere [41].

## B. MBE-grown surfaces

In parallel with cleaved surfaces, we have also studied properties of layers grown by MBE. In order to ascertain surface termination of the latter, we deposited slightly off-stoichiometric epitaxial layers. The typical topographic features and STS spectra are shown in Fig. 4. Image 4(a) corresponds to layer grown with Ti flux exceeding that of Sr. The surface is covered by unit cell high steps and linear features roughly perpendicular to the steps. Identical "nano-line" structure was reported on titanium rich surfaces [42,43]. The SrO-terminated surface shown in Fig. 4(c) was prepared by growing in Sr-rich condition, similar to what reported by Nie *et al*. [44]. Such surface is dominated by stepped-terrace structure without any linear feature as in Fig. 4(a). Slightly curve step edges were observed along with surface morphology that appears as a connection of small islands. These features indicate a layer-by-layer growth with incomplete terrace formation [45]. Conductance spectra, averaged across the surfaces, for these two surfaces are shown in Fig. 4(b) and (d), respectively. Spectral characteristics of the MBE-grown surface terminations were found to be similar to those of cleaved surfaces (Fig. 3). The noise level for these spectra is higher than those acquired on the cleaved surfaces, due to different acquisition electronics used in the two experiments. The similarity of the spectra between the cleaved and MBE-grown surfaces reaffirms our association of the respective surface terminations for the latter.

Notably, there are no observable in-gap states, other than the oxygen vacancy peak on SrO-terminated surface, on MBE surfaces. In particular, the tails of disorder-induced states that extend into the band gap from both the VB and CB in the spectra of Figs. 2 and 3 are absent in Fig. 4. We attribute this difference to the better (flatter) morphology of grown surfaces. Additionally, we note the absence in the spectra of Fig. 4 of any signature of the VB edge, i.e. expected to occur near -2 V (as in Figs. 2 and 3). This apparent lack of band edge is a common feature in STS studies of large bandgap materials [46]. It generally signifies band bending during the STS measurement, which can occur either due to the electric field between the tip and surface extending into the sample or due to surface charging by tunnel current. The second effect can lead to significant changes in the apparent tunneling barrier height, i.e. changes in the observed values for κ (as reported in Section III(A)), and is likely the main effect in the present work. In any case, with an increase in the density of in-gap states, band bending by both effects is suppressed. Hence, e.g., in the spectrum of Fig. 3(b) for the TiO$_2$ termination after annealing, there are many more disorder-induced states and the VB edge becomes much more apparent after the annealing. For the case of the spectra in Fig. 4, their lack of apparent VB edge is completely consistent with their lack of in-gap states, i.e. at negative sample voltages the surfaces become positively charged, and there are relatively few in-gap states available to inhibit the concomitant band bending of the SrTiO$_3$ due to that surface charge.

Returning to the prominent in-gap state seen in Fig. 4(d) for the SrO-terminated surface, its intensity is somewhat lower than for the cleaved surfaces (after annealing), which we attribute simply to a lower density of oxygen vacancies since the grown sample was not as strongly reduced as the cleaved samples. As already mentioned, the absence of other in-gap states indicates a higher surface quality, i.e. with a lower density of disorder-induced surface states, compared to the cleaved ones. However, as seen in Fig. 4(c), the surface of this sample still appears somewhat rough. Hence, a sample with flatter growth surface was prepared and studied, as shown in Fig. 5.

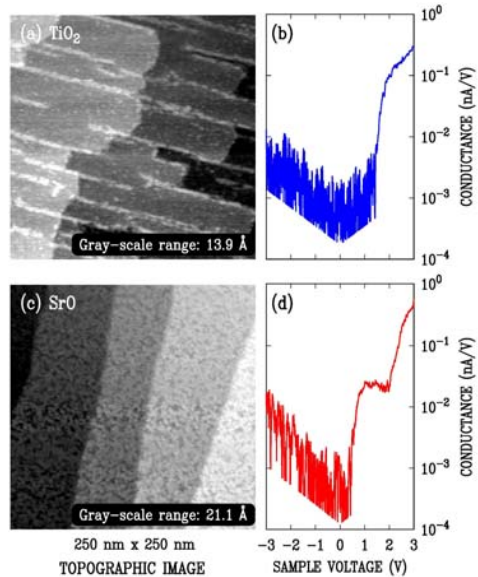

**FIG. 4** (Color online) Topographic images obtained with the sample bias of +1.5 V and tunnel current of 0.5 nA, and conductance spectra of MBE-grown (a), (c) TiO$_2$-terminated surface and (b), (d) SrO-terminated surface. The noise level for these spectra is clearly apparent at low conductance values (the noise level varies with voltage, due to the normalization of the spectra to constant tip-sample separation). Conductance values less than about one order-of-magnitude below the noise level are not shown.

The STM image of Fig. 5(a) reveals a significantly flatter surface (still with SrO termination) than that of Fig. 4(c) with each terrace completely filled. Considering the spectra of Fig. 5(b), we see that with diminishing surface disorder, the oxygen vacancy peak shifts towards the Fermi level (0 V), and an additional peak appears on a negative side of the spectrum. The position of these two peaks was found to vary slightly across the surface, as illustrated in Fig. 5(b). The emergence of the peak on a negative voltage side can be well explained by considering the influence of compensating acceptor-like states on the surfaces, originated from the surface disorder, as discussed in Section IV.

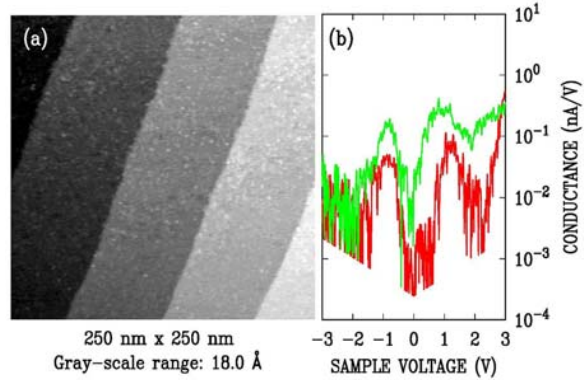

**FIG. 5** (Color online) (a) topographic image acquired at a sample bias of +1.5 V and tunnel current of 0.5 nA, and (b) conductance spectra acquired from two nearby locations on SrO-terminated surface.

## IV. BAND BENDING MODEL

According to results of Guisinger *et al.* [32] as well as ones presented here, the spectra clearly reveal a Fermi level that is located within the band gap even though the substrate is heavily doped with donors, i.e. niobium. Therefore, upwards band bending occurs in these n-type samples, with the Fermi level pinned near mid-gap. We interpret this band bending as arising from the presence of disorder-induced in-gap states that act to accept electrons donated from niobium donors, as illustrated in Fig. 6(a). We generally observe the in-gap feature that we have associated with oxygen vacancies to be located *above* the Fermi level, although on the flattest surfaces (with fewest disorder-induced state) we find these states to straddle the Fermi level, Fig. 6(b). Our interpretation is presented in Fig. 6. For a relatively high density of disorder-induced states relatively to the vacancy states, the Fermi level is constrained by the former, and ends up below the vacancy states. However, when the density of disorder-induced states is sufficiently reduced, then these have a smaller influence on determining the Fermi level position. In that case, the band bending is reduced, and the states of the vacancies approach and/or cross the Fermi level. This behavior, as exemplified in the spectra of Fig. 5, is indicative of *donor* character of the vacancy states. (In the absence of any disorder-induced states whatsoever, then the Fermi level would lie *above* the vacancy donor states, although we have not fully achieved that situation for the surface of Fig. 5).

Based on this model, the two peaks observed in the spectra Fig. 5(b) are actually associated with the same band of donor states, i.e. oxygen vacancy states. A distinct minimum in conductance at the Fermi energy may be attributed to effects such as Coulombic interaction or Mott hopping within a partially filled impurity band [47–50]. This movement of the Fermi level toward the conduction band as the surface becomes flatter serves as a solid proof for the existence of both the acceptor-like disorder-induced states and the donor-like vacancy states. Nevertheless, this explanation does not provide a clarification for the absence of in-gap state for TiO$_2$-terminated surfaces; such a state arising from oxygen vacancies has in fact been previously observed on TiO$_2$-terminated surfaces by photoemission spectroscopy (PES) [23]. In order to further analyze this situation, we have performed electronic ground-state calculations with different charge states of the oxygen vacancy; we now turn to a discussion of those results.

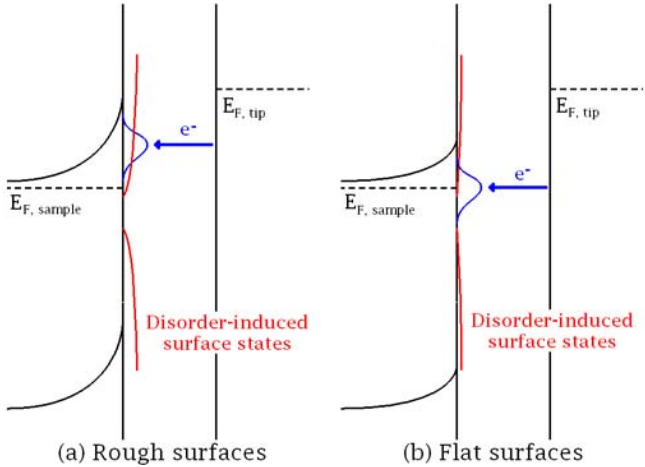

**FIG. 6** (Color online) Schematic diagram illustrating different Fermi pinning position for (a) rough surfaces and (b) flat surfaces. The density of disorder-induced states is lower on the flatter surfaces, as illustrated in the figure, and hence the amount of band bending is reduced on the flatter surfaces.

## V. CALCULATIONAL METHODOLOGY

Prior studies have shown the relevance of oxygen vacancies to observed in-gap states [51]. However, the experimentally observed gap-feature must be accompanied by a charge transition level from the oxygen vacancy for it to be correctly associated with theoretical calculation. In this work, we examine SrTiO$_3$ (001) slabs with one oxygen vacancy per simulation cell. We analyze the relative energetics of an oxygen vacancy, in various charge states, as a function of its position relative to the surface. The electronic ground-state calculations for the neutral ($V_O^0$), +1 charged ($V_O^+$) and +2 charged ($V_O^{+2}$) oxygen vacancies were performed using DFT with the local spin density approximation (LSDA+U) for exchange and correlation as implemented in the Quantum Espresso simulation package [52]. To account for strong electronic correlations we use a Hubbard U term [53] and a spin polarized calculation was employed because of the magnetic nature of the oxygen vacancies [54]. Our results reported here utilize U = 5 eV for Ti d states, although the qualitative trends in our results (e.g. resonant state for TiO$_2$ termination vs. in-gap state for SrO termination) are consistent with values of U in the range 4 to 5 eV [55,56]. We employ ultra-soft pseudopotentials [57] including semicore electrons for O (2$s$2$p$), Sr (4$s$4$p$5$s$) and Ti (3$s$3$p$4$s$3$d$). For each slab a 2×2 in-plane periodicity and 4 SrTiO$_3$ layers along the z-direction was used, along with a vacuum region of ~15 Å. A cutoff energy of 80 Ry and a Monkhorst-Pack special k-point mesh of 4×4×1 for the Brillouin zone integration was found to be sufficient to obtain better than 10 meV/atom convergence. Structural optimizations were performed by fixing the in-plane lattice constant of one SrTiO$_3$ unit to that of the theoretical bulk SrTiO$_3$ lattice constant ($a_0$ = 3.85 Å). All ions were then relaxed until the Hellmann-Feynman forces were less than 10 meV/Å.

The formation energy was calculated using [58]

$$E^f\left[V_O^q\right] = E_{tot}\left[V_O^q\right] - E_{tot}\left[SrTiO_3\right] + n_o\mu_o + q\left[E_F\right] \quad (1)$$

where $E_{tot}[V_O^q]$ is the total energy of supercell containing oxygen vacancy in a charge state $q$, $E_{tot}[SrTiO_3]$ is the total energy of a SrTiO$_3$ perfect crystal in the same supercell, and $\mu_O$ is the oxygen chemical potential. For a charged vacancy, the formation energy further depends on the Fermi level ($E_F$), which is the energy of the electron reservoir. Even with LDA+U, the band gap is underestimated and it needs to be scaled to the experimental value. While correcting the band gap, the formation energy obtained for a specific value of U also needs to be corrected. In this procedure, the formation energy of $V_O^{+2}$ is not affected as we vary U (i.e. change the band gap), since for $V_O^{+2}$ the Kohn-Sham gap state is empty and hence the total energy is unaffected as we vary both the band gap and the associated position of the Kohn-Sham gap state. For the $V_O^{+1}$ and $V_O^0$ cases, the formation energies are corrected assuming that the Kohn-Sham gap feature shifts with the CB, since the gap feature exhibits CB orbital character. Hence, we add $\left(E_{g,\exp} - E_{g,LDA+U(5)}\right) \cdot n$ to the formation energy, where $E_{g,\exp}$ is the experimental band gap, $E_{g,LDA+U(5)}$ is the band gap obtained from DFT calculation and $n$ the occupation of the Kohn-Sham gap state. To verify the accuracy of this correction and the choice of U, the transition levels thus obtained were evaluated for a bulk vacancy (Fig. 8(a)), yielding (+/++) and (0/+) levels located right at the CB minimum and 0.3 eV above the CB minimum, respectively. These results agree within a few tenths of an eV with those obtained by Janotti *et al.* [26], using a more accurate hybrid functional.

## VI. CALCULATIONAL RESULTS AND DISCUSSION

Figures 7(a) - (c) show the vacancy formation energies as a function of the Fermi level for the bulk, the SrO termination and the TiO$_2$ termination, respectively. For the SrO termination, we predict two transition levels, between +1 and +2 charge states (+/++), and between 0 and +1 charge states (0/+), when the Fermi level is 1.3 eV and 2.3 eV above the VB maximum, respectively. The position of the (0/+) level approximately matches the gap feature that we observed experimentally on the SrO termination. However, the lower (+/++) level was not observed in our experiments. The disorder-induced states on the surface would likely have pinned the Fermi level in between the two levels, such that only (0/+) level is empty but the (+/++) level is filled. In that case, the absence of (+/++) level can be attributed to a limited transport capability for in-gap states below the Fermi level of n-type material [46]. For in-gap surface states above the Fermi level (positive voltages), electrons tunneling into the states can tunnel through the depletion region into CB states, and observable current is thus achieved. However, for in-gap surface states below the Fermi level (small or moderate negative voltages), there are no bulk states available for the carriers to tunnel into, and thus their conductance is poor. Only when the density of surface states is large enough to allow *lateral* transport across the surface can these states be observed [30]. An exception to this situation occurs for a defect band of states is pinned right at the Fermi level (as in Fig. 5(b)), in which case both thermal excitation within the band as well as tunneling into CB states for a bulk Fermi level that is slightly above the CB minimum can produce observable features at negative voltages.

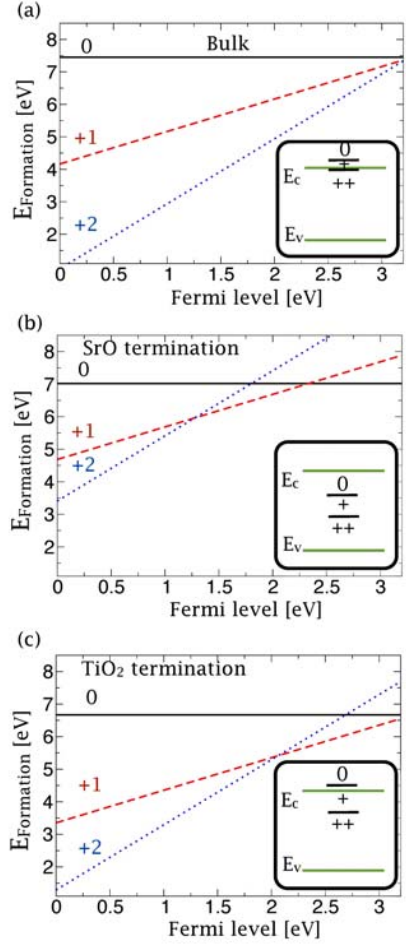

**FIG. 7**. (Color online) The formation energy as a function of Fermi level for different charge configurations for the oxygen vacancy in (a) the bulk, (b) the surface SrO layer, and (c) the surface TiO$_2$ layer. Insets show the resulting transition levels (Fermi level position at which transitions between charge states occur). For panel (c), the transition between +1 and 0 charge states occurs at a Fermi level position slightly above the CB minimum.

For TiO$_2$ termination, our calculation predicts a (+/++) level at 2.1 eV above the VB maximum, which in principle should be observable in the conductance spectra. However, no such discrete state was observed in the spectra. In some of our cleaved samples, we occasionally observe a weak, discrete feature in the upper half of the band gap for TiO$_2$-terminated surfaces after annealing. To further investigate the nature of the states on the different terminations, we compute the spin density of the in-gap state in its various charge states, as shown in Figs. 8(a)-(f). In the bulk, the oxygen ion has two nearest neighbor Ti ions. The wavefunctions of the vacancy in either $V_O^+$ or $V_O^0$ states are mostly made of Ti $3d$ orbitals pointing at the vacancy. On the SrO-terminated surface, the oxygen vacancy has only one Ti neighbor directly underneath. Therefore, the in-gap state is mostly made up of $d_{z^2}$ orbitals, which point towards the vacancy as clearly seen in Fig. 8(a) - (b). For the case of $V_O$ and $V_O^+$ at the TiO$_2$ surface, the orbital character is dominated by the $d_{(x^2-y^2)}$ and $d_{zy}$ orbitals pointing towards the vacancy, as shown in Fig. 8(d) - (e). This difference in orbital characteristic for the oxygen vacancy state at different terminations has a direct consequence for the sensitivity of the STS. The tunnel current is more sensitive to an orbital which points out in the direction perpendicular to the surface, since it has greater overlap with the wavefunctions of the tip. Therefore, it should be easier to detect the oxygen vacancy states on the SrO-terminated surface due to their dominant out-of-plane $d_z^2$ orbital characteristic. Detecting the oxygen vacancy states on $TiO_2$-terminated surface, on the other hand, is relatively difficult because the wave functions extend mostly along the surface. This characteristic of the wave function provides an explanation for the absence of any discrete in-gap state for the $TiO_2$ termination in our experiments.

Concerning the predicted (0/+) level on the $TiO_2$- terminated surface, in sharp contrast to the SrO-terminated case, it appears as a resonant level in the conduction band. Such a resonant level will autoionize, with the electron transferred to the conduction band. The resulting positively charged vacancies will cause downward band bending, leading to the formation of a 2DEG. This is the mechanism responsible for 2DEG formation on $SrTiO_3$ surface, as elucidated in some prior publications [59–65]. In contrast, for our surfaces produced experimentally, there apparently is always a sufficient number of disorder-induced states to accept electrons from the oxygen vacancies and thereby inhibit for 2DEG formation.

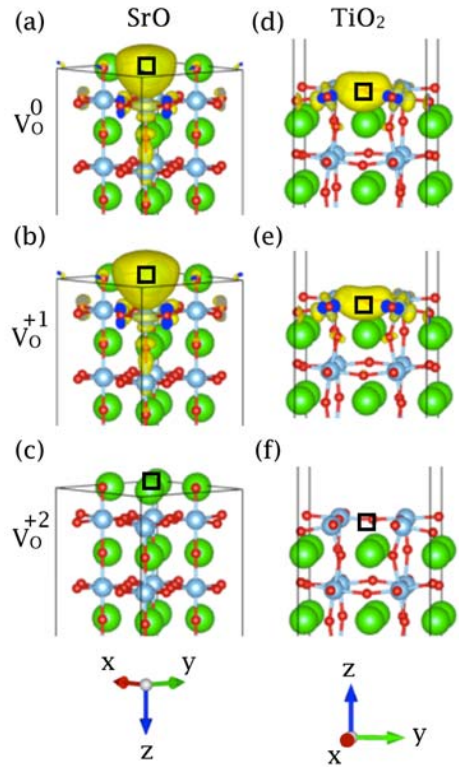

**FIG. 8.** (Color online) The majority spin density for the $V_O^0$, $V_O^+$ and $V_O^{+2}$ in the SrO and $TiO_2$ surface layer. The isosurfaces (yellow lobes) correspond to 2% of the maximum value in each plot. Green, blue and red balls represent Sr, Ti and O atoms, respectively. Square solid box represents position of the oxygen vacancy.

Regarding our observed position of the (0/+) on the SrO termination, the breadth of the spectral feature is quite large indicating a possible electron-phonon coupling. To evaluate such coupling, we calculated configuration coordinate (CC) diagrams together with the square of the vibronic (harmonic oscillator) wavefunction for the $V_O^0$ vacancy and the $V_O^+$ vacancy for the SrO termination. The actual transition level lies at the energy where the $V_O^+$ vacancy together with two electrons has the same energy as the $V_O^0$ vacancy. However, the peak in conductance spectra corresponds to the energy level where there is maximal overlap between the $V_O^+$ and $V_O^0$ vibronic states. With this vibronic coupling taken into account, the actual position of the surface oxygen vacancy transition level is found to lie 0.2 eV below the observed peak in the conductance spectra. Thus, for the observed oxygen vacancy peak shown in Fig. 3(b) positioned at 2.9 eV above the VB maximum, we estimate an actual transition level at 2.7 eV above the VB maximum. This value is reasonably close to the 2.3 eV transition energy found theoretically for the (0/+) transition level of the SrO-terminated surface (Fig. 7(b)).

To summarize, we find theoretically that a transition level above the CB edge is formed by vacancies in the outermost plane of $TiO_2$-terminated $SrTiO_3$ (001) surfaces (this result is essentially the same as believed to occur for vacancies in bulk $SrTiO_3$) [26]. This resonant level will produce a 2DEG, so long as compensating acceptor levels are not present on the surface (or in the bulk). In-gap levels, on the other hand, are produced by vacancies on either surface termination (and they also form for vacancies in the bulk, i.e. as the second donor level, when polaronic effects are included) [26]. The in-gap spectral feature commonly observed using PES [23,66], a technique which has a large probing area and a finite probing depth (~20 Å), likely is formed by a combination of these surface and bulk states.

## VII. SUMMARY

In summary, we have observed the single donor transition level of the surface oxygen vacancy on SrO-terminated $SrTiO_3$(001) by scanning tunneling spectroscopy. Segregation of bulk oxygen vacancies onto the room-temperature-cleaved surface gives rise to a large peak in the conductance spectra. Exposure of 10 Langmuir of molecular oxygen drastically reduces peak intensity, confirming the association with oxygen vacancies. The position of this peak was found to shift toward the Fermi level when the amount of surface disorder is reduced, as in the case of MBE-grown surfaces. Taking into account vibronic coupling, we determine a transition level at 2.7 eV above the valence band edge. The $TiO_2$-terminated terraces, on the other hand, did not exhibit any discrete in-gap state, which is attributed to the in-plane orbital characteristic of the oxygen vacancy state for these terraces. To understand the observed spectra, LSDA+U calculations were performed. Our calculated transition levels for a bulk oxygen vacancy match with the levels reported in Ref. [26]. Our predicted in-gap (double donor) and resonant (donor) levels for the $TiO_2$-terminated surface also agree with prior experimental observations, with the former level contributing to the in-gap feature observed by PES [21,23,25,38] and the latter level responsible for the reported formation of a 2DEG on that surface [18–20]. For the case of SrO-terminated surfaces, according to our calculations, our observed peak in the conductance spectra arises from a band of (0/+) levels. This band likely also contributes to the previously observed in-gap PES peak.

## ACKNOWLEDGMENTS

We thank Hemant Dixit and Valentino R. Cooper for useful discussions, and Mohamed Abdelmoula and Ying Lu for their help with titanium deposition. This research was supported by AFOSR Grant No. FA9550-12-1-0479, and it used resources of the National Energy Research Scientific Computing Center, supported by the Office of Science, US Department of Energy under Contract No. DEAC02-05CH11231.